\documentclass[aps,prd,groupaddress,onecolumn,usenatbib,nofootinbib,showpacs,altaffilletter]{revtex4}
\usepackage{graphicx}
\usepackage{dcolumn}
\usepackage{bm}
\usepackage{amssymb,amsmath,latexsym,amsfonts}
\usepackage[none]{hyphenat}

\begin{document}

\title{Scalar field as a Bose--Einstein condensate in a Schwarzschild-de Sitter spacetime}

\author{El\'ias Castellanos}
\email{ecastellanos@mctp.mx} 
\affiliation {Mesoamerican Centre for Theoretical Physics, \\Universidad Aut\'onoma de Chiapas.
Ciudad Universitaria, Carretera Zapata Km. 4, Real del Bosque (Ter\'an), 29040, Tuxtla Guti\'errez, Chiapas, M\'exico.}

\author{Celia Escamilla-Rivera}
\email{celia.escamilla@cosmo-ufes.org}
\affiliation {Mesoamerican Centre for Theoretical Physics, \\Universidad Aut\'onoma de Chiapas.
Ciudad Universitaria, Carretera Zapata Km. 4, Real del Bosque (Ter\'an), 29040, Tuxtla Guti\'errez, Chiapas, M\'exico.}

\author{Claus L\"ammerzahl}
\email{laemmerzahl@zarm.uni-bremen.de}
\affiliation {ZARM, Universit\"at Bremen, \\Am Fallturm, 28359 Bremen Germany.}

\author{Alfredo Mac\'{\i}as}
\email{amac@xanum.uam.mx}
\affiliation {Departamento de F\'{\i}sica, Universidad Aut\'onoma Metropolitana--Iztapalapa, \\PO.Box 55-534, Mexico D.F. 09340, M\'exico.}

\begin{abstract}
In this paper we analyze some properties of a scalar field configuration, where it is considered as a trapped Bose--Einstein condensate in a Schwarzschild--de Sitter background spacetime. In a natural way, the geometry of the curved spacetime provides an effective trapping potential for the scalar field configuration. This allows us to explore some thermodynamical properties of the system. Additionally, the curved geometry of the spacetime also induces a position dependent self--interaction parameter, which can be interpreted as a kind of \emph{gravitational Feshbach resonance}, that could affect the stability of the \emph{cloud} and could be used to obtain information about the interactions among the components of the system.
\end{abstract}

\maketitle

\section{Introduction}
\label{sec:intro}

Scalar fields always appear in many areas and situations in physics. In fact, one of the most remarkable ideas suggested in recent years is the concept of scalar field dark matter that was proposed as an alternative candidate to describe dark matter \cite{DM,DM1,DM2,DM3,DM4}.
In this model, the dark matter particle is assumed as a 0--spin boson, and this opens up the possibility of weakly interacting massive particles or WIMP's, axions, etc. This proposal also opens the door to the existence of scalar field dark matter as a Bose-Einstein condensate \cite{TonaRev,VSA,JB,KB,LP}.

 Furthermore, there are physical models related to the above ideas in the literature, such as the so--called \emph{hairy wigs models}, which basically describes the dynamical behavior of scalar field configurations surrounding black holes \cite{xxx}. The dynamics of the scalar field for this type of models can be analyzed from at least two different points of view. The scalar field can be treated as a test particle/field in a given gravitational background, i.e., feels gravity but its own gravity is neglected. Conversely, the system can be analyzed also from the \emph{self--gravitating point of view}, i.e., taking into account the gravity contributions produced by the scalar field itself.

Clearly, one of the main objectives associated with these systems is the possibility to form stable structure, in order to make them plausible candidates to describe dark matter galactic halos. In other words, these scalar field configurations must be stable and survive in the presence of black holes for enough time, of the order of cosmological scales. Let us mention that, according to the results exposed in \cite{xxx,QU,pelucas} both type of systems seems to be stable and can survive enough time in order to form structure. Consequently,  scalar fields could be relevant in the dark matter component present in the universe \cite{VSA,Barranco11,ure,ure1}.

On the other hand, the idea suggesting that scalar field configurations can be interpreted as relativistic Bose--Einstein condensates has already a long history \cite{dol,wei,eli,eli0,deLlano,NJP}. Nevertheless, the interpretation of the scalar field properties as a Bose--Einstein condensate still has to be fully understood, i.e., since only non--relativistic Bose--Einstein condensates have been created in the laboratory, the non--relativistic formalism associated with these systems agreed quite well with the experiments/observations. In this sense, the analysis 
of the properties related to a relativistic condensate are, at least until now, out of the reach of the current technology. However, from the theoretical point of view, it seems that relativistic Bose--Einstein condensates can be well understood as was analyzed in references \cite{dol,wei,eli,eli0,deLlano,NJP}, among others. Even more, the possible existence of scalar field dark particles, due to its bosonic nature, opens the possibility for the formation of cosmological Bose-Einstein condensates \cite{DM,BECs1,BECs2}. In other words, these scalar fields can be interpreted as a wave function that describes the evolution of a cosmological Bose--Einstein condensate together with some scalar potential that includes the possible self-interaction among the scalar field dark matter particles. It seems that there is a close-fitting relationship between the scalar field description of dark matter and Bose-Einstein condensates.

However, it not apropriate to call quasi bound distributions as cosmological condensates, as long as such distributions are described by a classical field without any reference to particles or quantum states. Even though the quasi bound distributions satisfy a very similar equation to the one satisfied by the stationary Bose-Einstein condensate, they have a very different origin, describe very different phenomena and are conceptually different. The similarity between the equations which describes each case was analyzed for example in \cite{rus}.

Recently, in \cite{NOS} we have investigated some existing analogies between the Klein--Gordon equation, which governs the dynamics of the scalar field, and the Gross--Pitaevskii equation, which drives the dynamics of a Bose--Einstein condensate. In \cite{NOS} it was shown that the Gross-Pitaevskii equation and the Klein--Gordon one present the same formal solutions (solitonic solutions) for a 1-dim Bose--Einstein condensate as well as for a scalar field \\configuration, both systems confined in a box. Additionally, it is remarkable that when we considered these systems in a curved background spacetime,  the scalar field configuration under study can be described through a Gross--Pitaevskii--like equation. We should notice, that the introduction of a curved spacetime induces, in a natural way a trapping potential that constrains the system, together with a position--dependent self--interaction parameter. The properties of the Gross--Pitaevskii--like equation obtained in \cite{NOS}, reinforce our guess that these type of scalar field configurations can be interpreted as Bose--Einstein condensates. The similarity works better in static and stationary situations where the role played by the time coordinate is not relevant. However,  this is a non trivial issue and clearly deserves deeper investigation.

Let us mention here that in the case of scalar field configurations surrounding black holes (\emph{or hairy wigs models}), the system can be described also from the Bose--Einstein condensate point of view \cite{NOS}. In such an scenario, it can be shown that a Gross-Pitaevskii equation emerges from the corresponding scalar field equation which describes the system.
In other words, even in this situation it is possible to analyze the dynamics of the system as a Bose--Einstein condensate in order to extract information. According to \cite{NOS}, the analogy between Bose--Einstein condensates and (test--)scalar field configurations surrounding black holes is remarkable.

However, deeper investigation is needed to support the description of these systems as dark matter and, clearly, the predictions obtained in \cite{NOS} must be confronted with observations in order to validate the model as a possible dark matter candidate.

In this work we perform a numerical analysis of some properties of a scalar field setup, where the scalar field is modeled as a trapped Bose--Einstein condensate in a fixed curved Schwarzschild-de Sitter background spacetime. The use of this metric could be relevant, in future work, to describe galactic halos as scalar fields like Bose-Einstein condensates or configurations where a black hole can be located at the center of a specific galaxy.

Moreover, we consider the scalar field as a test particle/field in the aforementioned background geometry. We analyze the scalar field \-configu\-ra\-tion from the point of view of Bose--Einstein condensates theory. We calculate the condensation temperature of the system for the \\ non--interacting case and we will give some insights related to the condensation temperature in the interacting case, which is a non--trivial issue too. Additionally, we assume that a non-relativistic bosonic gas is trapped in the potential induced by the curvature itself. Finally, we deduce some properties of the \emph{cloud} through the so--called Thomas--Fermi approximation which can be used to infer the size of the system, the density of particles and the total number of particles that may be constrained, in principle, by cosmological observations. The results obtained in this work allow us to conclude that there is a close relationship between scalar field configurations and Bose-Einstein condensates. Therefore, according to our proposal these systems seems to be a good candidate to describe dark matter in more realistic scenarios. This is clearly one of the most important modern issues which deserves further exploration.

\section{Curved Space Time and Induced Trapping Potential revisited}

We wish to study the analogy between the Klein--Gordon and the Gross-Pitaevskii equations in a curved spacetime. We start by summarizing some results already reported in \cite{NOS}.  First of all, in this work we consider the scalar field as a test particle/field in a curved background geometry i.e., the scalar field does not gravitationally back-react on the metric.
If a classical scalar field is considered as a test particle/field, it only
feels gravity, therefore its own gravity is neglected. Its dynamics \textbf{is} determined by a Klein--Gordon equation in a curved background spacetime:

\begin{equation}
\big[g^{\mu\,\nu}\,\nabla_\mu\,\nabla_\nu-\left(\sigma^2 + \lambda\,\rho_n\right)\,\big]\Phi=0, \label{eq:KG1}
\end{equation}
where $\sigma$ is the corresponding mass parameter and $\lambda$ is the parameter that quantify the strength of interactions among the bosonic particles within the system.
Also, the number density $\rho_n$ is defined as usual:
\begin{equation}
\rho_n=\Phi^*\,\Phi. \label{eq:rhon}
\end{equation}
Here, it is important to remark that the field $\Phi$ is a classical function not an operator, and can be interpreted as the \emph{macroscopical} wave function of the system or the \emph{order parameter} as in standard theory of condensates. Therefore, $\rho_n$ is well defined as the number density in this scenario.

Let us consider a spherically--symmetric--static background spacetime of the form
\begin{equation}
ds^2=-F(r)\,c^2\,d\,t^2 + \frac{dr^2}{F(r)} + r^2\,d\Omega^2, \label{eq:ele}
\end{equation}
with $d\Omega^2=d\theta^2 +\sin^2\theta\,d\varphi^2$, and $c$ the speed of
light in vacuum. By solving the vacuum Einstein field equations including a cosmological constant one can determine the explicit form of the metric structural function $F$.

We proceed by considering the dynamics of a scalar test field, $\Phi$, with a scalar self--interacting potential given by
\begin{equation}
V(\Phi\,\Phi^*)=\frac{\sigma^2}{2}\,\Phi^*\,\Phi +\frac{\lambda}{4}\,[\Phi^*\,\Phi]^2,\label{eq:V}
\end{equation}

In order to obtain the time independent form of the Klein--Gordon equation, Eq.\,(\ref{eq:KG1}), we use the monopolar component of the scalar field with harmonic time dependence:
\begin{equation}
\Phi=e^{i\,\omega\,t}\,\frac{u(r)}{r}.\label{eq:phiu}
\end{equation}
Thus, the Klein--Gordon equation, Eq.\,(\ref{eq:KG1}) reduces to the
following Gross--Pitaevskii--like equation
\begin{equation}
\Bigg(-\frac{d^2\,}{d\,{r^*}^2} + V_{\rm eff} + \lambda_{eff}\,\rho_n\Bigg)\,u=\mu_{eff}\,u.
\label{eq:KGGP}
\end{equation}
Where we have introduced the $r^*$ coordinate
\begin{equation}
r^*=\int\,\frac{d\,r}{F}. \label{eq:r_es}
\end{equation}
Consequently, the effective trapping potential can be written as
\begin{equation}
V_{\rm eff}(r)=F\,\,\left(\sigma^2 + \frac{F'}{r} \right), \quad r=r(r^{*}),
\label{eq:Veff}
\end{equation}
where the effective self--interaction parameter can be defined by
\begin{equation}
\lambda_{eff}=\lambda\,F.
\label{lameff}
\end{equation}
Additionally, we have an effective chemical potential $\mu_{eff}=\omega^2/c^2$.

In order to obtain stationary (or quasi stationary) solutions for the scalar field, the
curvature of the spacetime itself should confine the scalar field and the confinement region can be represented by the effective potential (\ref{eq:Veff}). In other words, the gravitational background is able to enclose the scalar field.

It is important to stress the fact that from Eq.\,(\ref{eq:KGGP}) the curvature of the spacetime induces also a spatially dependent
self--interaction, through the parameter $\lambda F$. This \emph{effective} self--interaction
parameter $\lambda_{eff}=\lambda F$ (position dependent) could be interpreted as a
kind of \emph{gravitational Feshbach resonance} induced by the curvature of
spacetime, and could affect for instance, the stability of the system, as it happens in usual
condensates \cite{pethick2002,nat}.

Let us consider a Schwarzschild black hole within a de Sitter spacetime, the metric coefficient $F$ of Eq.~(\ref{eq:ele})
reduces to
\begin{equation}
F=1 - \frac{2\,M\,G}{c^2\,r} - \frac{\Lambda}{3}\,r^2, \label{eq:FSdS}
\end{equation}
where $M$ is the black hole mass and $\Lambda$ is the cosmological constant.

Choosing a mass--scale $M_0$, and a distance--scale $R_0$, we construct the dimensionless quantity $q=G\,M_0/c^2\,R_0$, then the black hole mass
is a factor times the mass--scale, i.e., $M=n\,M_0$. Similarly the distance is a multiple of the distance--scale
$r=x\,R_0$, with $n, x$ dimensionless constants. The scale parameters $M_{0}$ and $R_{0}$ are arbitrary and depend on various physical parameters of the model, e.g., the mass of the black hole, the mass of the scalar particle, the radius of the galactic halos, etc. These are chosen in such a way that the dimensionless quantities are convenient for the numerical analysis.
Since $\Lambda$ has units of curvature, that
is, inverse of area, we construct the dimensionless quantity $\nu= \Lambda\,{R_0}^2/3$, so the
metric coefficient $F$ given in Eq.\,(\ref{eq:FSdS}) reduces to the following dimensionless form
\begin{equation}
F=1 - 2\,q\,\Bigl(\frac{n}{x}\Bigr) - \nu\,x^2. \label{eq:FSdS1}
\end{equation}
\begin{figure}[!htb]
\centering
\includegraphics[width=0.45\textwidth,origin=c,angle=0]{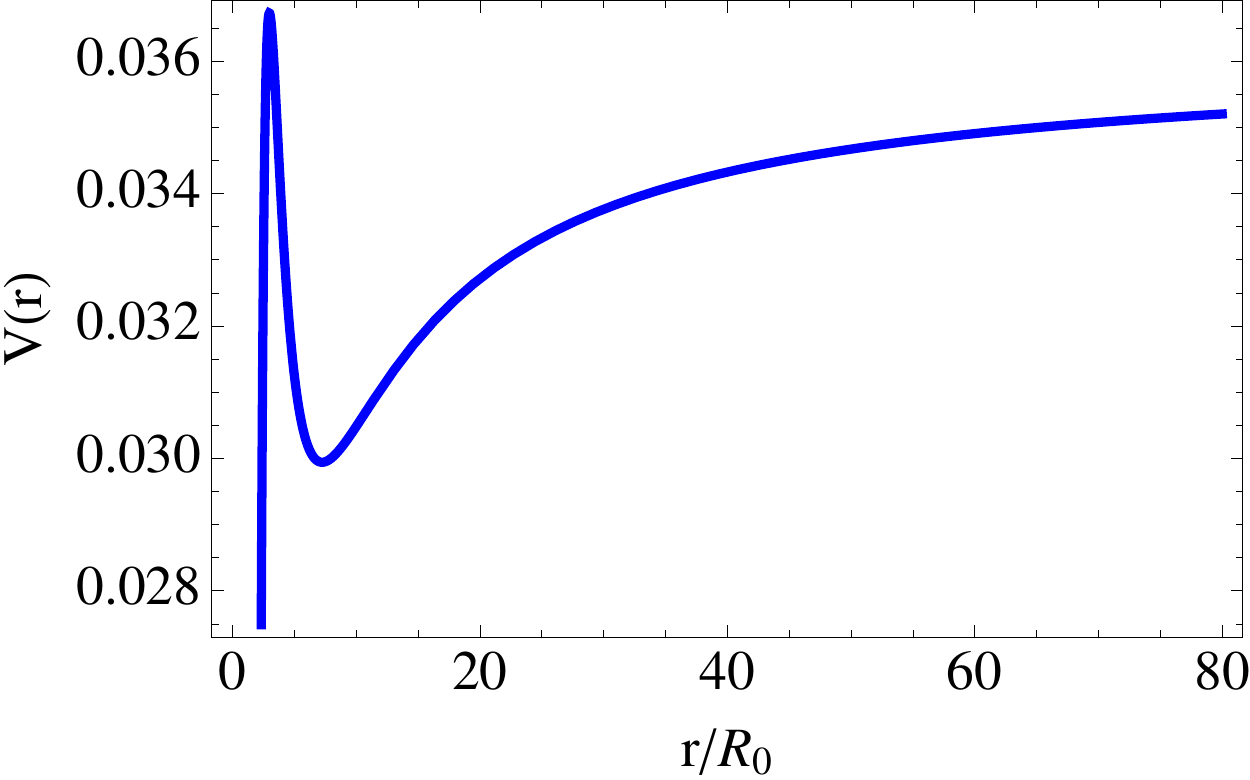}
\includegraphics[width=0.45\textwidth,origin=c,angle=0]{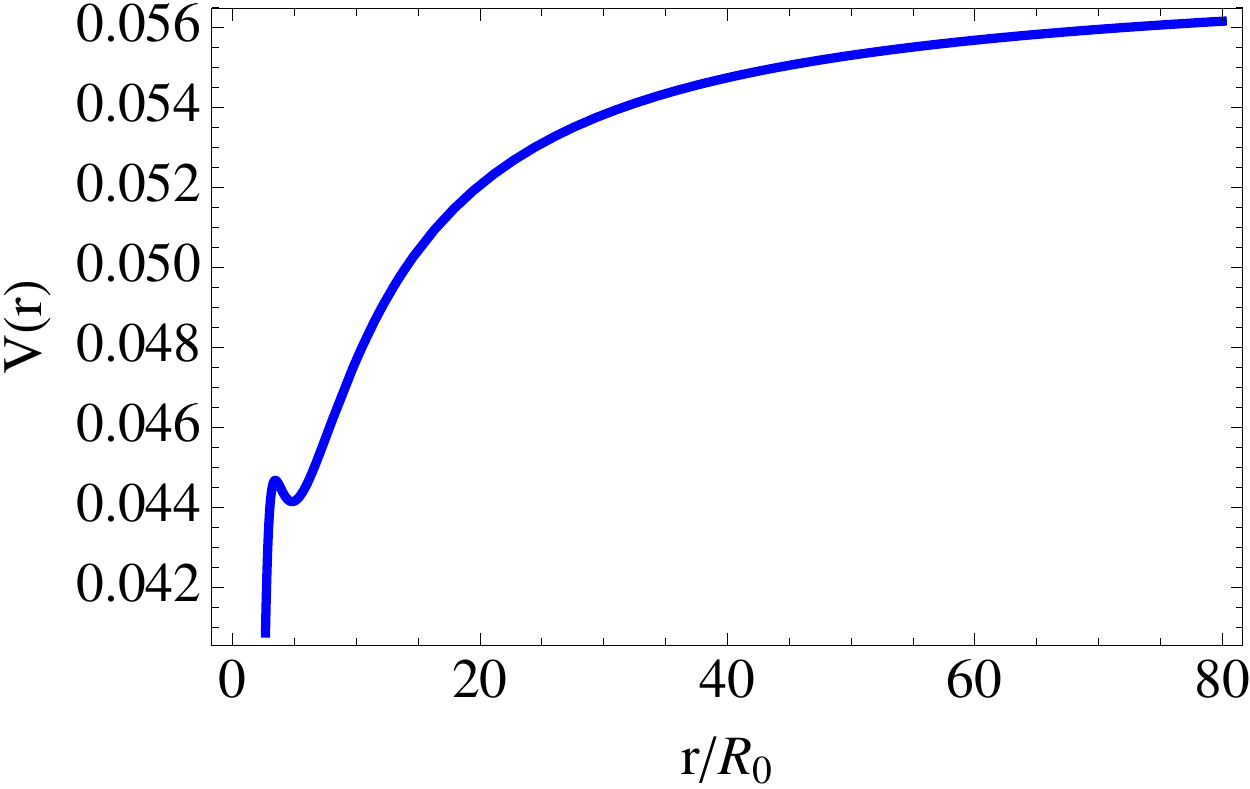}
\includegraphics[width=0.49\textwidth,origin=c,angle=0]{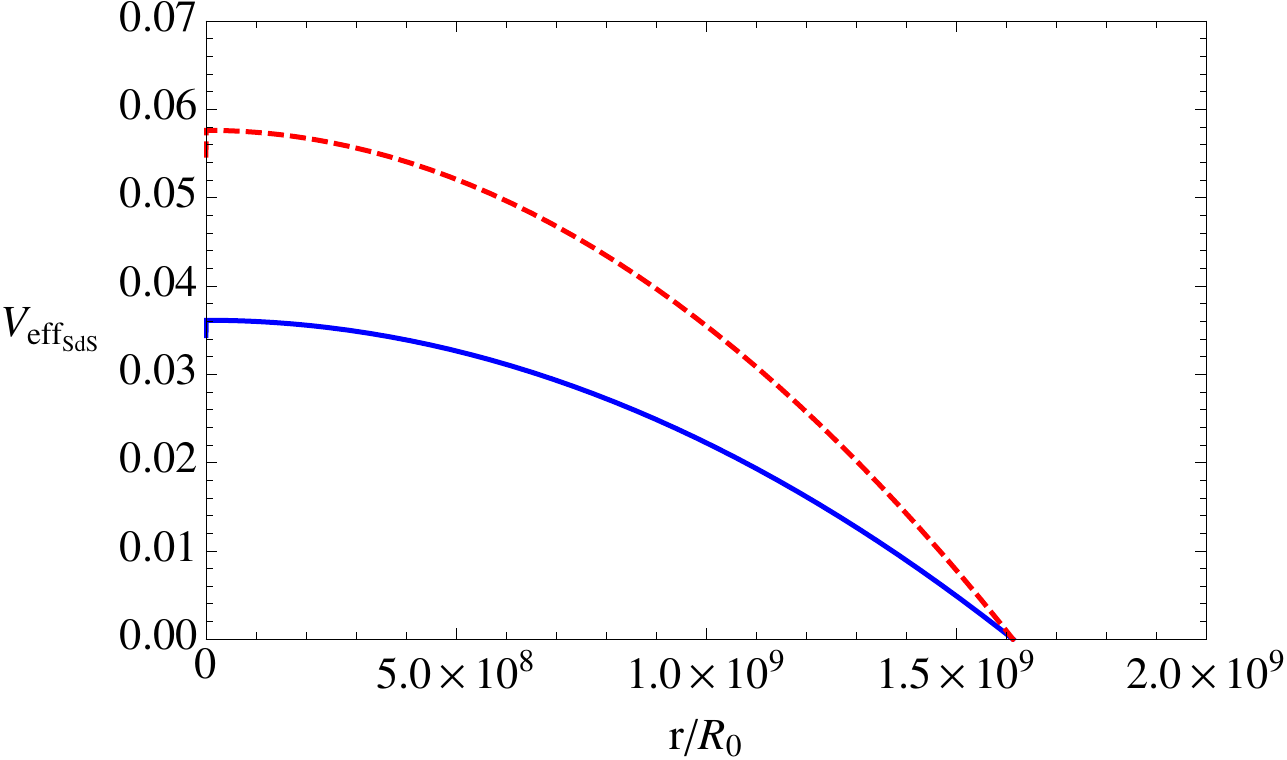}
\caption{\label{fig:case_sigmas_pot} \textit{Top left}: Effective potential Eq.\,(\ref{eq:Veff_SdS}) for the case $\sigma_1 =0.19$. \textit{Top right}: Effective potential   Eq.\,(\ref{eq:Veff_SdS}) for the case $\sigma_2 =0.24$. \textit{Bottom middle:} Cosmological Horizon for the Schwarzschild--de Sitter spacetime for the mass parameters $\sigma_1 =0.19$ (Blue solid line)  and $\sigma_2 =0.24$ (Red dashed line). We used here the following numerical values: $R_{0}=n=q=c=\hbar=1$ and $\nu=3.84 \times 10^{-19}$.}
\end{figure}

Now, the effective potential, Eq.\,(\ref{eq:Veff}) for this spacetime geometry reads
\begin{equation}
V_{\rm eff_{SdS}}=\frac{1}{{R_0}^2}\,\left(\alpha^2 - 2\,\nu + \frac{2\,q\,n}{x^3} \right)
\left(1 - \frac{2\,q\,n}{x} - \nu\,x^2\right),\label{eq:Veff_SdS}
\end{equation}
with $\alpha=R_0\,\sigma=R_{0}\,m c / \hbar$. For the scalar potential we use the same scale--distance as
the one used so that the parameter $\alpha$ is dimensionless. For $\alpha<\sqrt{2\,\nu}$, the
asymptotic behavior of the effective potential is positive, and for $x=2\,q\,n$ one has a characteristic black hole barrier.
Thus, we expect to have regions where there could exist bound states of the scalar field distribution \cite{NOS}.

For larger radius the effective potential grows and then starts to decline, reaching zero value at the cosmological horizon. The results are displayed in Figure~\ref{fig:case_sigmas_pot}. Thereby, the gravitational field generated by the black hole mass and the cosmological constant $\Lambda$ is endowed with trapped regions for the test scalar field.

In the Schwarzschild--de Sitter scenario, one deals with two kinds of horizons: one
related to the black hole and the other one associated with $\Lambda$ and the visible universe \cite{SdSC}.
For example, the effective potential Eq.(\ref{eq:Veff_SdS}) goes to zero when we consider $x=5386.37$, setting the size of the universe to $R=5.38\,\times 10^3\,{\rm Mpc}$ when a black hole is not present. However, when we consider a massive black hole, the internal horizon grows towards the external one in such a way that, for a value of n = 1036.6, the two horizons merge in one, and we obtain the so--called critical Schwarzschild--de Sitter spacetime.

In \cite{Barranco12} the evolution of a massive scalar field surrounding a Schwarzschild black hole was presented and configurations were found which can survive for arbitrarily long times when the black hole or the scalar field mass is small enough. While in \cite{Barranco13}, these
quasibound modes are discussed in the context of scalar field dark matter models. These facts support strongly the analogy between the scalar field and the order parameter describing Bose--Einstein condensates also in stationary curved backgrounds, since there are quasi--stationary distributions of the scalar field around the black hole, which
behave like a Bose--Einstein condensate. The election of these two scenarios is quite simple. The shape of the induced trapping potential seems to be an optimal option in which the geometry of the spacetime confines more remarkably the scalar field configuration (see Figure \ref{fig:case_sigmas_pot}).

From Fig.\,\ref{fig:case_sigmas_pot} we notice that the effective potential is able to enclose the quasi--stationary configuration of the scalar field. Thus, the analogy between the scalar field in a curved background satisfying Eq.\,(\ref{eq:KGGP}), and the order parameter satisfying the usual Gross--Pitaevskii stationary equation, is remarkable.

\section{Thomas-Fermi approximation and some properties of the cloud}

A formal solution for the scalar field distribution in a curved spacetime can be obtained within the so--called Thomas-Fermi approximation. This approximation is valid for systems at very low temperatures $T\ll T_{c}$ when the system is weakly interacting for sufficient large clouds and the kinetic energy is negligible with respect to the potential one. In this scenario the form of the Gross--Pitaevskii--like  Eq.\,(\ref{eq:KGGP}) reads:
\begin{equation}
\big(V_{\rm eff_{SdS}} + \lambda_{\rm eff_{SdS}}\,\rho_n\big)\,u=\mu_{eff}\,u\, ,
\end{equation}
with a solution given by
\begin{equation}
\rho_{n}=\frac{\frac{\omega^{2}}{c^{2}}-\frac{1}{{R_0}^2}\,\left(\alpha^2 - 2\,\nu + \frac{2\,q\,n}{x^3} \right)\,
\left(1 - \frac{2\,q\,n}{x} - \nu\,x^2\right)}{\lambda [1 - 2\,q\,\Bigl(\frac{n}{x}\Bigr) - \nu\,x^2]}. \label{eq:rhonTF}
\end{equation}

The size (boundary) of the cloud can be calculated from the expression
\begin{equation}
\label{size}
\frac{\omega^{2}}{c^{2}}=\frac{1}{{R_0}^2}\,\left(\alpha^2 - 2\,\nu + \frac{2\,q\,n}{x^3} \right)\left(1 - \frac{2\,q\,n}{x} - \nu\,x^2\right),
\end{equation}
which fixes the size of the system within this approximation.

Moreover, the number of particles within the cloud, i.e., within the condensate, is given by the normalization condition
\begin{eqnarray}
N=\int{\rho_{n} r^2 dr}, \label{eq:density}
\end{eqnarray}
where
\begin{eqnarray}
\rho_{n} = \frac{\frac{w_0^2}{c^2}-\frac{\left(-\frac{2 n q R_0}{r}-\frac{\nu  r^2}{R_0^2}+1\right) \left(-2
   \nu +\frac{2 n q R_0^3}{r^3}+\sigma ^2\right)}{R_0^2}}{\lambda  \left(-\frac{2 n q
   R_0}{r}-\frac{\nu  r^2}{R_0^2}+1\right)}.
\end{eqnarray}

The effective potential with our new definitions for Schwarzschild--de Sitter case (\ref{eq:Veff_SdS}) reads
\begin{eqnarray}
V_{\rm eff_{SdS}}&=&\frac{1}{{R_0}^2} \left[\alpha^2 -2\nu
 +\frac{2qn}{(r/R_0)^3}\right]
  \left[1-\frac{2qn}{(r/R_0)} -\nu(r/R_0)^2\right].
\label{eq:Veff_SdS1}
\end{eqnarray}

Let us compute the roots for $r$ using Eq.\,(\ref{size}). The corresponding roots with physical meaning are given by:
\begin{eqnarray}
r_{1}= 2.24,\quad
r_{2}= 1.21 \times 10^{9}. \quad
\label{eq:erres}
\end{eqnarray}
These roots fix the size of the \emph{cloud}, i.e., dimensionless radius up to $\sim 10^{9}$.
In the above results we used the numerical values:
$ \sigma_1 = 0.19,\,
\sigma_2 = 0.24,\,
R_0 = 1, \,
n= 1,\,
q= 1,\,
\nu = 3.84 \times 10^{-19},\,
w_0 = 10,\,
c=1,\,
\lambda = 0.01,\,
\hbar = 1.$
These roots determine also the integration limits of the normalization condition Eq.\,(\ref{eq:density}). The numerical integration of Eq.\,(\ref{eq:density}) gives us as a result $N=1.24 \times 10^{26}$ particles for the $\sigma_1$ case and $N=1.06 \times 10^{27}$ particles for the $\sigma_2$ case. Also with these values the density $\rho_{n}$ Eq.(\ref{eq:rhonTF}) is positive and well defined.

Notice that within the Thomas--Fermi approximation the numerical values obtained above are only valid near the minimum of the effective potential which contains the \emph{cloud}. Furthermore, we consider ultra--light scalar field with mass around $m_\Phi\,c^2 \equiv 10^{-24}\,{\rm eV}$, and for $\hbar\,\sigma/c = m_\Phi$, we obtain that the corresponding parameter $\sigma$ for such ultra-light scalar mass is $5.06\times10^{-18}\,{\rm m}^{-1}$ in ordinary units. Within this approximation the total mass of the \emph{cloud} for $10^{26}$ particles with a mass of order $m_{\Phi}\approx10^{-60 }$\,kg leads to a total mass of order $N m_{\Phi}\approx10^{-34}$kg, these values are negligible when compared with the total mass of the system. In other words, the use of the \emph{test field approximation} is justified.

In order to extract information concerning the density of the \emph{cloud} viewed as a Bose--Einstein condensate in comparison to the critical density of the universe we can set for the $\sigma_2$ case the following cosmological parameter: $\Omega=\rho_{n}/\rho_{c}=2.1/\rho_c (\text{M}_\odot \text{Mpc}^{3})$, where we assume that basically from (\ref{eq:erres}) the size of the \emph{cloud} is of the order $10^{9}$. Then with
$\rho_c =3H^{2}_{0}/8\pi G $, $H_0 =67.81\pm 0.92$ and $G=4.3\times 10^{-6}$ Kpc $\text{km}^2\text{s}^{-2}\text{M}_\odot^{-1}$ \cite{Ade:2015xua} we can provide a fraction of the matter density of the \textit{cloud} that could be extrapolated as an example of dark matter for a particular astrophysical system.

\begin{figure}[!htb]
\centering
\includegraphics[width=0.49\textwidth,origin=c,angle=0]{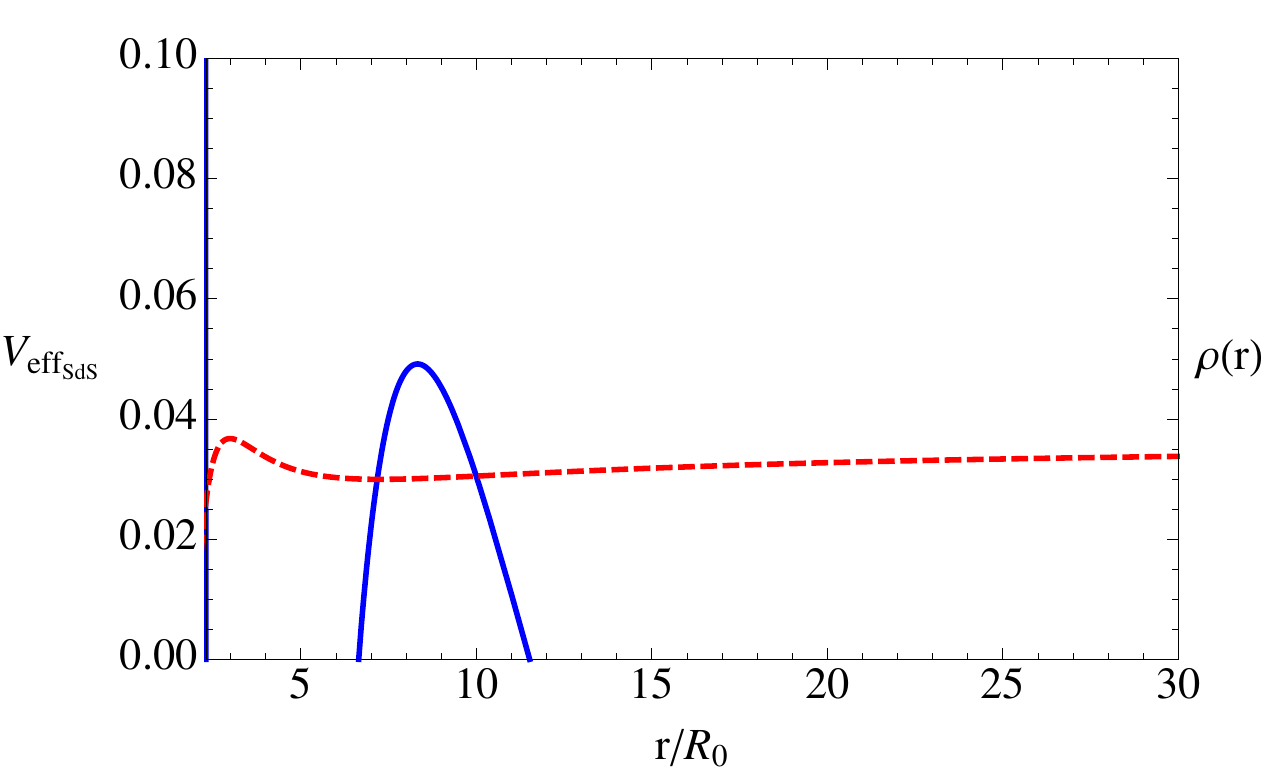}
\includegraphics[width=0.49\textwidth,origin=c,angle=0]{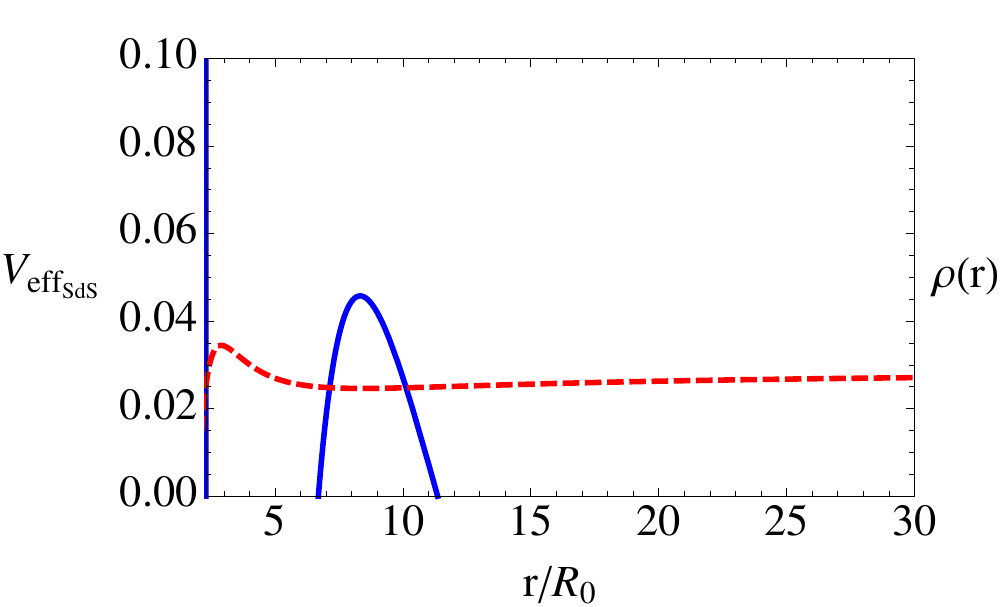}
\caption{\label{fig:case_sigmas} \textit{Left}: Effective potential (red dashed line) Eq.\,(\ref{eq:Veff_SdS}) and density solution (blue solid line) Eq.\,(\ref{eq:rhonTF}) for the case $\sigma_1 =0.19$. \textit{Right}: Effective potential (red dashed line) Eq.\,(\ref{eq:Veff_SdS}) and density solution (blue solid line) Eq.\,(\ref{eq:rhonTF}) for the case $\sigma_1 =0.24$. Notice that the Thomas--Fermi approximation shows that the maximum of the density is located at the minimum of the effective potential for both values of the mass parameter. This indicates that the Thomas--Fermi approximation leads to well defined values of the effective potentials and their corresponding densities.}
\end{figure}

\begin{figure}[!htb]
\centering
\includegraphics[width=0.49\textwidth,origin=c,angle=0]{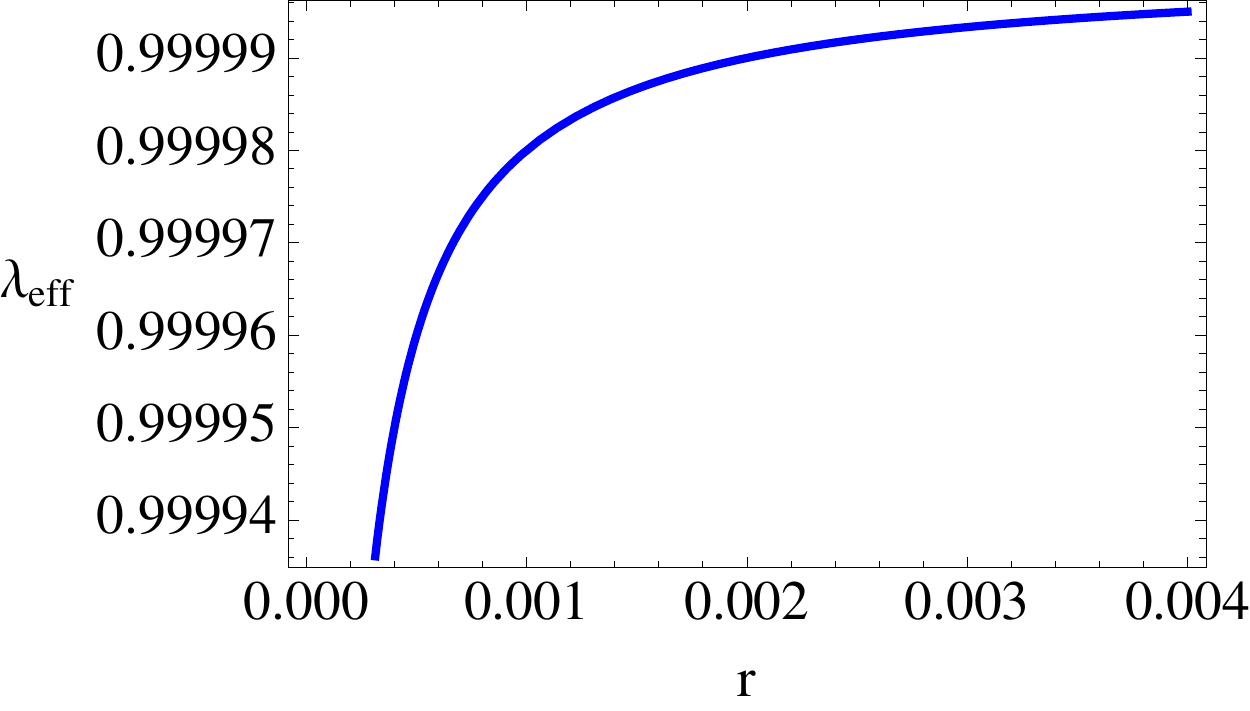}
\includegraphics[width=0.499\textwidth,origin=c,angle=0]{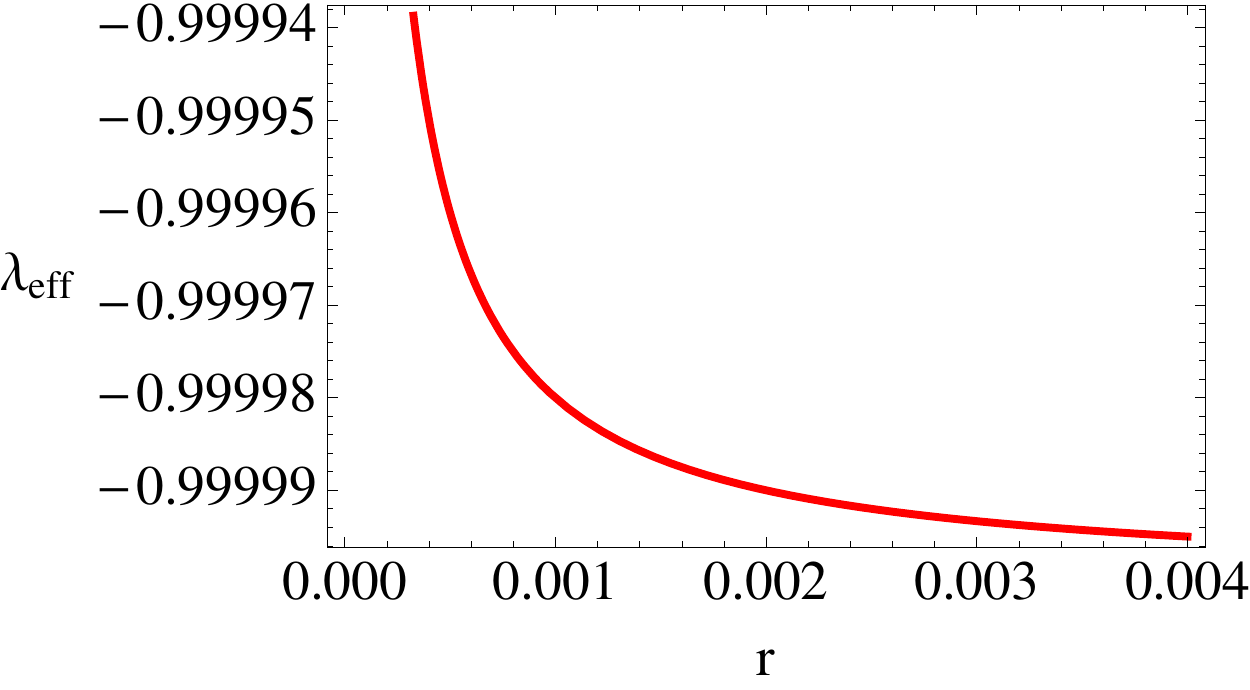}
\caption{\label{fig:case_lambdas} \textit{Left}: Effective self--interaction parameter $\lambda_{\rm eff_{SdS}}$ for the case with $\sigma_1 =0.19$. \textit{Right}: $\lambda_{\rm eff_{SdS}}$ for the case with $\sigma_2 =0.24$. These results are obtained for $\lambda>0$.}
\end{figure}

Notice that $\lambda_{\rm eff_{SdS}}=\lambda (-\frac{2 n q R_0}{r}-\frac{\nu  r^2}{R_0^2}+1)$ is negative for $\lambda>0$ and tends to
$-\infty$ when $r <\sqrt{-\frac{3}{2\Lambda}+\sqrt{(\frac{3}{2\Lambda})^{2}+6\frac{GM}{\Lambda c^{2}}}}$, because the dependence $2\,M\,G/c^2\,r$, see Figure\,\ref{fig:case_lambdas} on the left side. In other words, large changes in $\lambda_{\rm eff_{SdS}}$ can be produced by small changes in the coordinates.

\begin{figure}[!htb]
\centering
\includegraphics[width=0.44\textwidth,origin=c,angle=0]{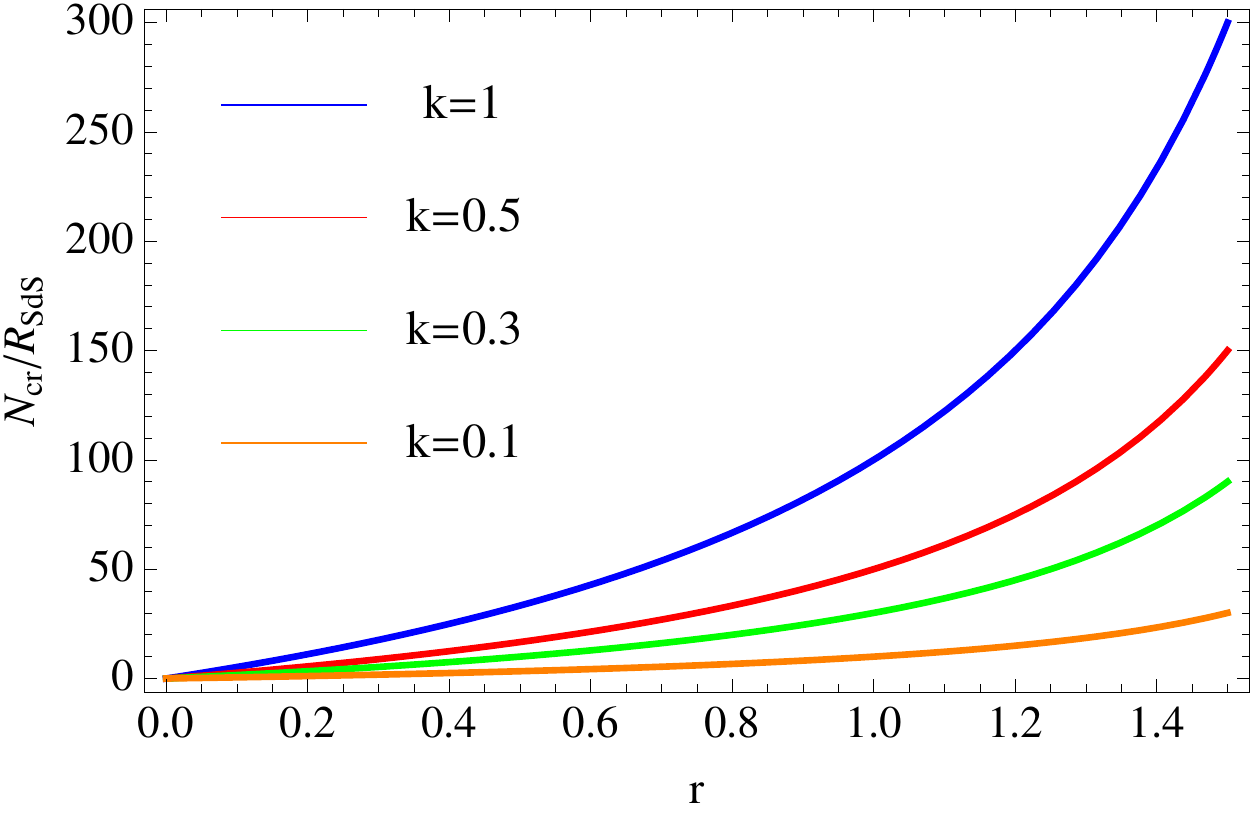}
\includegraphics[width=0.48\textwidth,origin=c,angle=0]{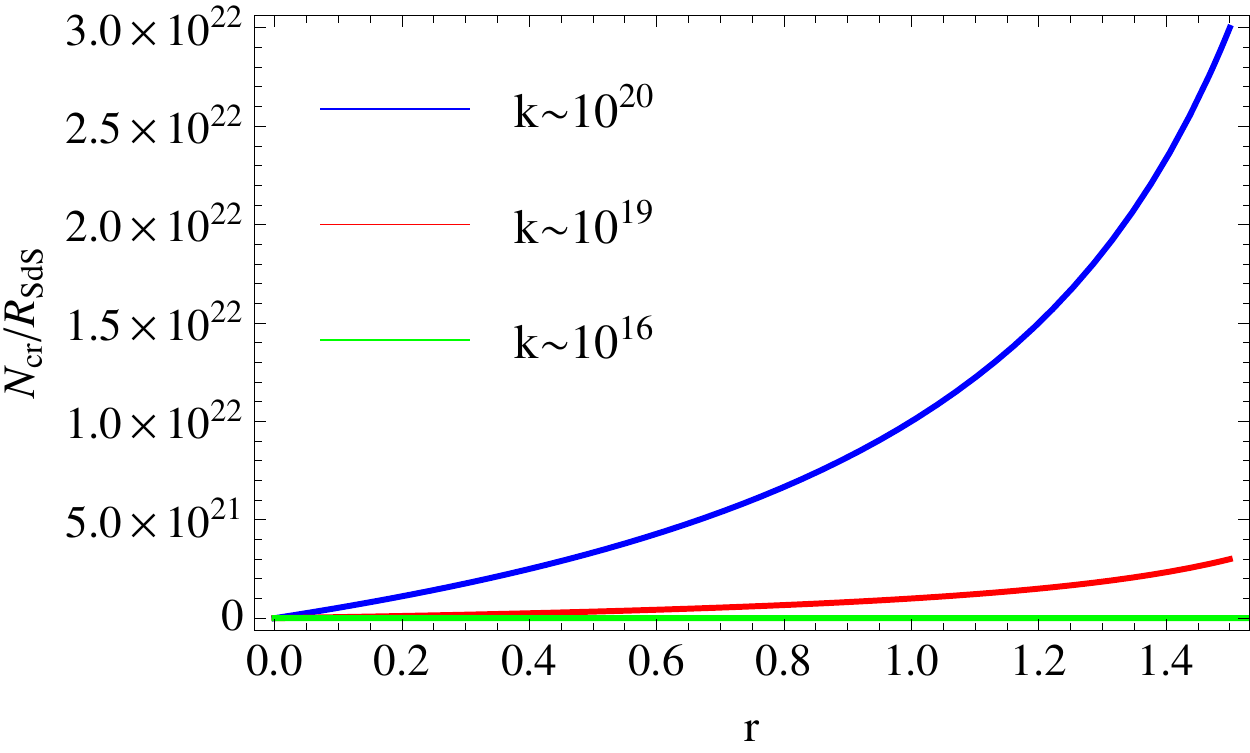}
\caption{\label{fig:case_Ncr_k} \textit{Left}: Critical number of the particle parameter Eq.(\ref{eq:Ncr}) for a stable scenario. \textit{Right}: Critical number of the particle parameter Eq.(\ref{eq:Ncr}) for an unstable scenario.}
\end{figure}

In this scenario we are able to define a critical number of particles related to the stability of the system according to \cite{nat} as follows
\begin{equation}
N_{cr}=k\frac{R_{SdS}}{|\lambda_{\rm eff_{SdS}}|}, \label{eq:Ncr}
\end{equation}
where $k$ is the \emph{stability coefficient} \footnote{In usual Bose--Einstein condensates this parameter is a positive dimensionless constant. Also, its value depends on some properties of the magnetic trap and is related to the stability of the system see \cite{nat} for details.} and $R_{SdS}$ is the size of the \emph{cloud}. 
Moreover, $R_{SdS}$ is basically of the order of $10^{9}$ as it was inferred from the results Eq.(\ref{eq:erres}). If $N$ obtained from Eq.\,(\ref{eq:density}), which we assume as the number of particles in the condensed state, i.e., $10^{26}-10^{27}$ particles, is less than $N_{cr}$, then the system is stable for some values of the constant $k$. Otherwise it is unstable. These properties can be used to explore the stability of the cloud. In other words, this analysis suggests that when $N>N_{cr}$ then particles are lost from the system within this approximation. We notice in Figure \,\ref{fig:case_Ncr_k} that the system is stable for small values of the constant $k$ which implies that $N_{cr} \sim 10^{11}$ particles. Conversely, for large values of $k$ the system is unstable, since $N_{cr}\sim 10^{31}$. Bounds for the constant $k$ must be constrained using astrophysical data, in order to analyze the stability of the system and consequently to extract information about the possibility for these systems to form stable structures.

\bigskip

\section{Semiclassical approximation and condensation temperature}

When the time in not relevant, i.e. in the case of stationary/static spherically--symmetric backgrounds, we expect that the semiclassical approximation usually used to describe some properties of Bose--Einstein condensates can be more accurate for our system, which brings as a consequence the energy conservation of the \emph{cloud}. Therefore, the semiclassical single particle energy spectrum associated with our effective Gross--Pitaevskii Eq.\,(\ref{eq:KGGP}) reads
\begin{equation}
\epsilon=\frac{p^{2}}{2m_{\Phi}}+V_{\rm eff_{SdS}}+\lambda_{\rm eff_{SdS}}\rho_{_T},
\end{equation}
where $V_{\rm eff_{SdS}}$ is given by Eq.\,(\ref{eq:Veff_SdS1}) and $\lambda_{\rm eff_{SdS}}$ by Eq.\,(\ref{lameff}) with $F=1-2q\Bigl(\frac{n}{x}\Bigr)-\nu x^2$. Additionally, $\rho_{_T}$ is the corresponding density of particles above the condensation temperature.
Recall that the mass of the scalar field particle is around $m_\Phi\,c^2 \equiv 10^{-24}\,{\rm eV}$, and thus $\sigma=5.06\times10^{-18}\,{\rm m}^{-1}$. Here we have also assumed that the bosonic gas inside the effective potential $V_{\rm eff_{SdS}}$ is non relativistic.
The semiclassical single particle energy spectrum can be used to calculate and estimate some relevant properties associated with the system. Let us calculate the corresponding condensation temperature.

The condensation or transition temperature is defined as the lowest temperature at which the macroscopic occupation of the ground state appears.
In the semiclassical approximation, the single--particle phase--space distribution for bosons may be written as \cite{pethick2002,P}
\begin{equation}
\label{SPSD}
f(\epsilon)=\frac{1}{e^{\beta(\epsilon-\mu_{eff})}-1},
\end{equation}
where $\beta=1/\kappa T$, $\kappa$ is the Boltzmann constant, $T$ is the temperature, and $\mu_{eff}$ is the effective chemical potential defined above. Clearly, $\epsilon$ depicts the single particle energy spectrum.
Consequently, the total number of particles $N$ obeys the
normalization condition
\begin{equation}
 N=N_{0}+\frac{1}{(2 \pi \hbar)^{3} }\int d^{3}r\hspace{0.1cm} d^{3} p\hspace{0.1cm}f(\epsilon),
\label{NC}
\end{equation}
where $N_{0}$ is the number of particles in the ground state.

The spatial density above the condensation temperature is given by
\begin{equation}
\label{n1}
\rho_{T}= \rho_{0}+\int  d^{3}p\,f(\epsilon).
\end{equation}
where $\rho_{0}$ is the density of the ground state.

It is straightforward to show that the density is given by
\begin{equation}
\label{den1}
\rho_{_T}=\rho_{0}+\Lambda_{dB}^{-3}\,g_{3/2}(Z_{eff}).
\end{equation}
where $\Lambda_{dB} \equiv (2\pi \hbar ^{2}/m \kappa T)^{1/2}$, is the thermal wavelength and the function $g_{k}(*)$ is the so--called Bose--Einstein function \cite{P}.
In this scenario we have that the function $Z_{eff}$ can be interpreted as an effective fugacity. However, the effective fugacity depends on the effective external potential and also on the effective self--interaction parameter, together with the effective chemical potential through the following relation
\begin{equation}
\label{ZEFF}
Z_{eff}=\exp\,\beta\,[\mu_{eff}-V_{\rm eff_{SdS}}-\lambda_{\rm eff_{SdS}}\,\rho_{T}].
\end{equation}

In the non--interacting case, i.e., setting $\lambda_{eff}=0$, the effective fugacity is given by $Z_{eff}=\exp\,\beta\,[\mu_{eff}-V_{\rm eff_{SdS}}]$.

Let us analyze the case of Schwarzschild--de Sitter spacetime in this context.
Using the normalization condition Eq.\,(\ref{NC}), or equivalently, integrating the density Eq.\,(\ref{den1}) one obtains
\begin{eqnarray}
N&=&N_{0}+\frac{8\pi^{2}(2m\kappa T)^{3/2}}{(2\pi \hbar)^{3}} \, \sum_{l=1}^{\infty} \Bigg[\frac{\exp{(\beta \mu_{eff}})^{l}}{l^{3/2}}\,\int r^{2}\,dr\exp{(- V_{\rm eff_{SdS}}/\kappa T)^{l}}\Bigg].
\end{eqnarray}

Assuming also that the system lies in the thermodynamical limit, we can safely set $\mu_{eff}=0$ and $N_{0}=0$ in order to extract the condensation temperature from the above expression. Thus, we obtain
\begin{equation}
\label{NI1}
N=\frac{8\pi^{2}(2m\kappa T_{c})^{3/2}}{(2\pi \hbar)^{3}} \, \sum_{l=1}^{\infty} \Bigg[\frac{1}{l^{3/2}}\,\int r^{2}\,dr\exp{(- V_{\rm eff_{SdS}}/\kappa T_{c})^{l}}\Bigg],
\end{equation}
where $T_{c}$ is the condensation temperature in the non--interacting case. Equation (\ref{NI1}) must be solved numerically.
The integral to solve for $T_{c}$ is
\begin{equation}
\label{NI}
N=\frac{8\pi^{2}(2 \sigma_{s} T_{c})^{3/2}}{(2\pi)^{3}} \, \sum_{l=1}^{\infty} \Bigg[\frac{1}{l^{3/2}}\,\int^{r_f}_{r_i} r^{2}\,dr\,e^{-( V_{\rm eff_{SdS}}/ T_{c})^{l}}\Bigg],
\end{equation}
in units where $\kappa=1$, $\hbar=1$, and $s=1,2$ corresponds to the cases $\sigma_{1}$ and $\sigma_{2}$, respectively.
Additionally, we have inserted the roots obtained in expressions Eq.(\ref{eq:erres}), i.e., $r_{1} =r_{i}$ and $r_{2}=r_{f}$. Let us fix the minimal value for the number of particles $N$ using the result of the numerical integration of Eq.\,(\ref{eq:density}) which is  $N=1.24 \times 10^{26}$ particles for $\sigma_1$ case and $N=1.06 \times 10^{27}$ particles for $\sigma_2$ case.
Now, with this minimal value of $N$ we are able to calculate the condensation temperature for both $\sigma$ cases obtaining as a result $T_c \approx 5\times 10^{-4}$, where we have used the numerical values mentioned above. Notice that the condensation temperature $T_{c}$ obtained above is dimensionless.

\begin{figure}[!htb]
\centering
\includegraphics[width=0.44\textwidth,origin=c,angle=0]{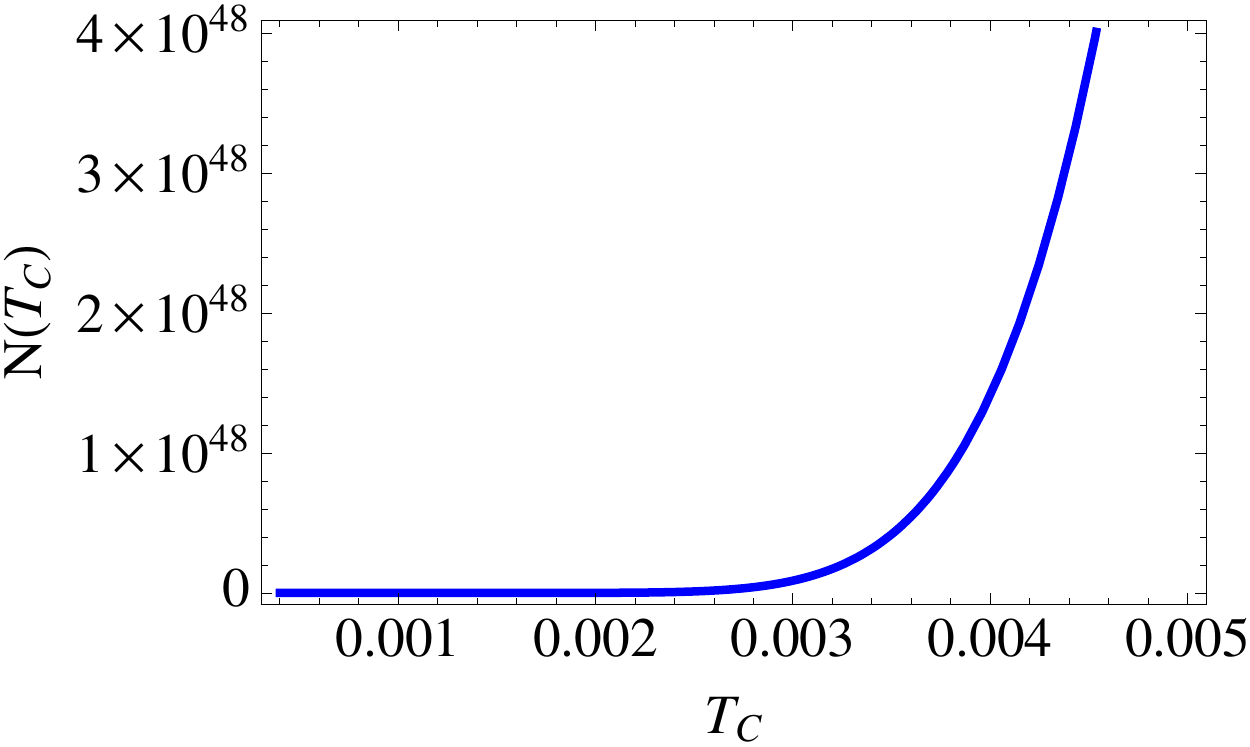}
\includegraphics[width=0.45\textwidth,origin=c,angle=0]{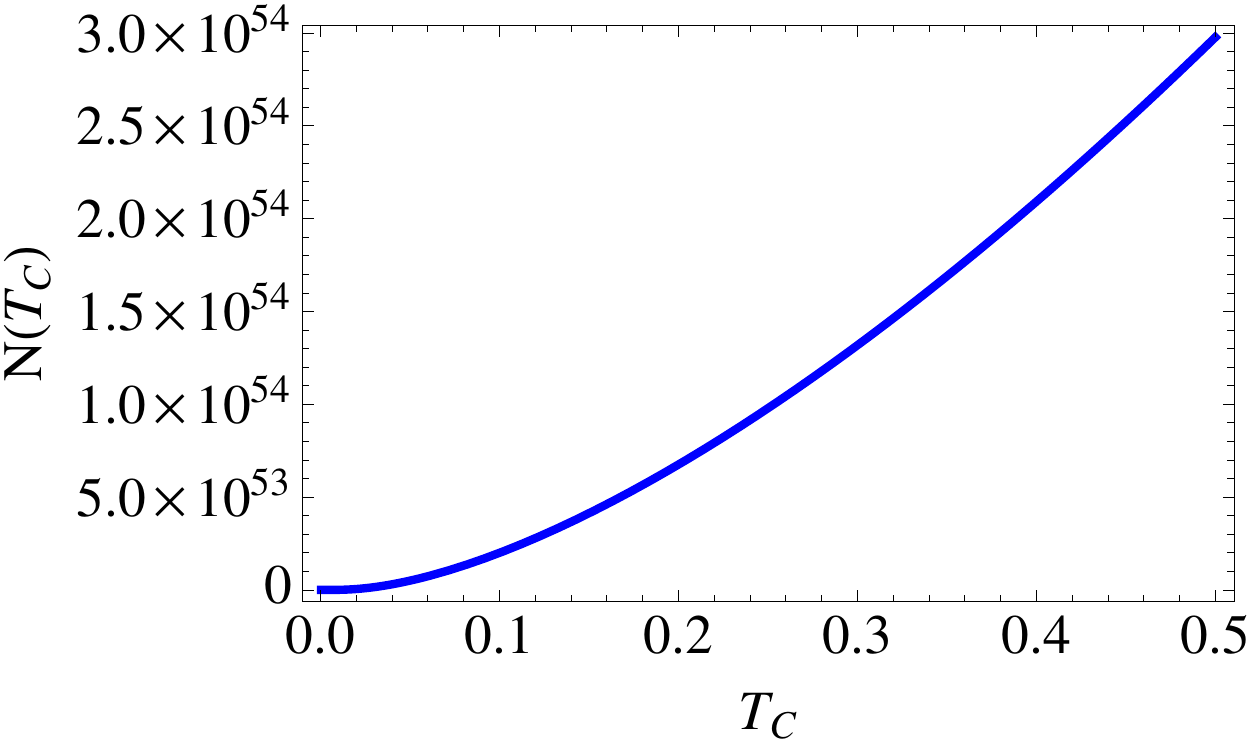}
\caption{\label{fig:numerical_solutions}
\textit{Left:} Evolution of $N$ Eq.\,(\ref{NI}) at small critical temperatures. \textit{Right:} Evolution of $N$ Eq.\,(\ref{NI}) at large critical temperatures.
There is not significant differences between $\sigma$-cases. }
\end{figure}

After this estimation, we proceed to compute the numerical solution of Eq.\,(\ref{NI}), i.e., the functional relation between the condensation temperature and the number of particles. In Figure \ref{fig:numerical_solutions} it is shown the behavior of the number of particles as a function of the condensation temperature, for the mentioned cases of interest.

It is worthwhile to stress the fact that the condensation temperature is an increasing function of the number of particles. In other words, large number of particles implies higher condensation temperatures. Conversely, small number of particles implies lower condensation temperatures.


\subsection{Comments on the condensation temperature in the interacting case}

So far, we have assumed that the effective fugacity is a solely function of $\beta$, $\mu_{eff}$ and
$V_{\rm eff_{SdS}}$. This assumption is not justified through an interacting process where $\lambda_{\rm eff_{SdS}}\neq 0$.
For that, the interacting case must be solved also numerically. In such a case the total number of particles above the condensation
temperature is now given by
\begin{eqnarray}
\label{INT}
N&=&N_{0}+\frac{8\pi^{2}(2m\kappa T)^{3/2}}{(2\pi \hbar)^{3}} \, \sum_{l=1}^{\infty}\Bigg[\frac{\exp{(\beta \mu_{eff}})^{l}}{l^{3/2}}\, \int^{r_f}_{r_i}  r^{2}\,dr\exp{\Bigl((- V_{\rm eff_{SdS}}-\lambda_{\rm eff_{SdS}}\,\rho_{T})/\kappa T\Bigr)^{l}}\Bigg].\,\,\,\,\,\,\,\,\,\,\,\,\,\,\,\,
\end{eqnarray}
The latter integral must be solved self--consistently. Additionally, at the condensation tempe-\\rature the number of particles in the ground state can be also neglected. However, the effective chemical potential cannot be set equal to zero due to interactions \cite{pethick2002}.  In usual condensates, at the condensation temperature $T_{c}$ for large $N$, in the mean field approach, the chemical potential can be obtained by evaluating the density at the center of the potential, i.e., at $r=0$, times the value of the self--interaction parameter which in this case is a constant, depending only on the corresponding s--wave scattering length.

In other words, this means that the critical density at the center of the system is approximately equal to the one of the uniform case, i.e., without trap. In our case the parameter $\lambda_{\rm eff_{SdS}}$, which accounts for the interaction among the constituents of the \emph{cloud}, depends on the position and the center of the effective potential $V_{\rm eff_{SdS}}$, which it is not situated at $r=0$. This last assertion
leads us to conclude that in order to calculate the corresponding $T_{c}$ in the interacting case, the effective chemical potential must be evaluated at the minimum of the induced trapping potential, through the corresponding density.

Additionally, Eqs.\,(\ref{NI}) and (\ref{INT}) can be used to obtain the corresponding shift in the condensation temperature with respect to the non--interacting case and consequently the condensed fraction of the system. This deduction can be used to analyze if the interactions within the \emph{cloud} could be representative and hence able to shift significantly the condensation temperature of the system from the non--interacting case. As a first step to study this scenario we need to solve Eq.\,(\ref{INT}) numerically. Such process can be quite tricky depending of the form (or behaviour) of $\lambda_{eff}$, at any rate, we already compute the $r$ range in which this numerical integration can be done. The second step to pursue will consist to find the adequate values of $k$, where this scenario experiences interesting change in the $N_{cr}/R_{SdS}$ ratio.
This analysis is non trivial and a more detailed study of this case will be reported elsewhere.

\section{Conclusions}

In this work, we have analyzed and described the behavior of a scalar field configuration in a Schwarzschild--de Sitter background spacetime, assuming as a fundamental fact that the scalar field configuration in this scenario can be interpreted as a trapped Bose--Einstein condensate. We have used the Thomas--Fermi approximation in order to obtain and analyze some properties of the scalar field configuration, i.e., \emph{the cloud}. We have calculated, within this approximation, the corresponding density of particles and the size of the system. With all this information we establish some limits in which our model could be more accurate. One of these scenarios suggests that the stability of the system can be analyzed according to \cite{nat} or at least within some analogue approach. However, all these results associated with the stability of the \emph{cloud} depend on the \emph{stability coefficient} $k$. This parameter must contain information about the trapping potential as in usual condensates, i.e., information concerning the curvature of the corresponding background under consideration.

We are able to estimate some possible values of the parameter $k$ in order to \emph{determine} the stability of the system. Let us point out that these values will change depending of the region of interest which lies in the functional form of the $\lambda_{\rm eff_{SdS}}$.
We emphasize that some astrophysical data could be useful in order to constrain such a constant together with the predictions of our model and consequently to test the validity of our approximations, for instance in dark matter models.

Also, we calculated the condensation temperature of the system, assuming a collection of non relativistic bosons trapped by the curvature of the geometry itself, and when the contributions of the interactions among the constituents of the system are not present. We remark that this temperature should depend on the astrophysical data for a specific configuration.

We made comments about the interacting case which we believe must change our predicted value of the condensation temperature. Since the corresponding condensation temperature must be shifted with respect to the non-interacting case, the presence of $\lambda_{\rm eff_{SdS}}$ should be relevant for the estimation of the condensation temperature in a realistic scenario.
Deeper investigation and confrontation with observations are needed to support the description of these Bose--Einstein condensates systems as a possible dark matter candidate and as a model for black hole accretion.
In recent years great attention has been devoted to the neutrino mixing phenomenon that allowed the development of other approaches to explain $\Lambda$CDM and dark energy \cite{ref1}, where on one hand the energy content of the neutrino mixing vacuum condensate \cite{ref2} can be interpreted as dynamically evolving dark energy that, at present epoch, assumes the behavior and the value of the observed cosmological constant. Additionally, a vacuum condensate due to neutrino and quark mixing in two regimes \cite{ref3}: (A) the regime of the matter dominated universe with adiabatic index $w=p/\rho$ ranging between 0 and 1/3; and (B) the present epoch regime, dark energy dominated universe, with $w \approx -1$ have been studied. Under reasonable boundary conditions the vacuum condensate from particle mixing can provide contributions to the dark energy compatible with the observed value of the cosmological constant, leading to dark energy values compatible with those inferred from observations.
On the other hand, some cosmological implications of ultraviolet quantum effects have led to a condensation of Born--Infeld matter\,\cite{ref4}. This invisible non-linear electrodynamics coupled to neutrinos can provide both a Dirac neutrino mass term and a dark energy contribution, i.e., an effective cosmological constant.


\acknowledgments

This work was supported by DFG-CONACyT Grant Nos.\,B330/418/11 and 211183, by CONA-\\CyT Grant No. 166041F3.
E.C. and C.E-R acknowledges MCTP for financial support. C.L. acknowledges support by the DFG Research Training Group 1620 \emph{Models of Gravity} and by the QUEST
Center of Excellence.



\end{document}